\documentclass[10pt, conference]{IEEEtran}
\IEEEoverridecommandlockouts
\usepackage{geometry}
\usepackage{amsmath}
\usepackage{amssymb}
\usepackage{bm}
\usepackage{siunitx}
\usepackage{mathtools}

\usepackage{xcolor}
\usepackage{graphicx}
\usepackage{epsfig}
\usepackage{tikz}
\usetikzlibrary{plotmarks,patterns,decorations.pathreplacing,backgrounds,calc,arrows,arrows.meta,spy,matrix}
\usepackage{tikzscale}

\usepackage{pgfplots}
\pgfplotsset{
    compat=1.3,
    legend style={font=\footnotesize, fill opacity=0.7,  draw opacity=1, text opacity=1, draw=white!15!black, legend cell align=left, align=left}, 
    width=0.8\columnwidth, 
    scale only axis,
    height=4cm,
    yminorticks=false,
    xminorticks=false,
    label style={font=\footnotesize},
    title style={font=\small},
    tick align=outside,
    tick pos=left,
    tick style={color=black},
    tick label style={font=\footnotesize},
    grid style={line width=.1pt, draw=gray!20},
    major grid style={line width=.1pt,draw=gray!20},
}
\usepgfplotslibrary{colorbrewer, groupplots, patchplots}
\usepackage[font=footnotesize]{subcaption}
\usepackage{url}
\usepackage[utf8]{inputenc}
\usepackage{comment}
\usepackage{enumitem}
\usepackage[colorlinks = true,
            linkcolor = black,
            urlcolor  = black,
            urlbordercolor=blue,
            citecolor = black,
            anchorcolor = black]{hyperref}
\usepackage{xurl}
\usepackage{colortbl}
\usepackage[capitalize]{cleveref}
\usepackage{pbox}
\usepackage[acronyms,nonumberlist,nopostdot,nomain,nogroupskip]{glossaries}
\newacronym{3gpp}{3GPP}{3rd Generation Partnership Project}
\newacronym{bs}{BS}{base station}
\newacronym{ppp}{PPP}{Poisson Point Process}
\newacronym{su}{SU}{scheduling unit}
\newacronym{ofdm}{OFDM}{Orthogonal Frequency Division Multiplexing}
\newacronym[shortplural=WuDs, longplural=devices equipped with wake-up receiver]{wud}{WuD}{device equipped with wake-up receiver}
\newacronym{wur}{WuR}{wake-up receiver}
\newacronym{wus}{WuS}{wake-up signal}
\newacronym{idwu}{IDWu}{identity-based wake-up}
\newacronym{fifo}{FIFO}{First In First Out}
\newacronym{awgn}{AWGN}{Additive White Gaussian Noise}
\newacronym{mac}{MAC}{Medium Access Control}
\newacronym{pdf}{pdf}{probability density function}
\newacronym{rb}{RB}{resource block}
\newacronym{ucwu}{UCWu}{UniCast Wake-up}
\newacronym{ul}{UL}{uplink}
\newacronym{dl}{DL}{downlink}
\newacronym{harq}{HARQ}{hybrid automatic repeat request}
\newacronym{iiot}{IIoT}{Industrial Internet of Things}
\newacronym{ack}{ACK}{Acknowledgement}
\newacronym{mmtc}{mMTC}{Massive Machine Type Communication}
\newacronym{urllc}{URLLC}{ultra-reliable low-latency communication}
\newacronym{pmf}{pmf}{probability mass function}
\newacronym{iot}{IoT}{Internet of Things}
\newacronym{cps}{CPS}{Cyber-Physical System}
\newacronym{nr}{NR}{New Radio}
\newacronym{rv}{r.v.}{random variable}
\newacronym{ss}{SS}{Synchronization Signal}
\usepackage[normalem]{ulem}

\usepackage{multirow}
\usepackage{tablefootnote}
\usepackage{booktabs}
\usepackage{tabularx}

\definecolor{amaranth}{rgb}{0.9, 0.17, 0.31}
\definecolor{steelblue}{RGB}{176,196,222}
\definecolor{darkblue}{RGB}{0,0,139}
\definecolor{lightblue}{RGB}{31,119,180}
\definecolor{deepskyblue}{RGB}{0,191,255}
\definecolor{lightskyblue}{RGB}{135,206,250}
\definecolor{lightgray}{rgb}{0.82, 0.82, 0.82}
\definecolor{gray}{RGB}{140,140,140}
\definecolor{darkgray204}{RGB}{204,204,204}
\definecolor{darkgray224}{RGB}{224,224,224}

\newcommand{\mc}[1]{\mathcal{#1}}   

\makeatletter
\def\@citex[#1]#2{\leavevmode
\let\@citea\@empty
\@cite{\@for\@citeb:=#2\do
{\@citea\def\@citea{,\penalty\@m\ }%
\edef\@citeb{\expandafter\@firstofone\@citeb\@empty}%
\if@filesw\immediate\write\@auxout{\string\citation{\@citeb}}\fi
\@ifundefined{b@\@citeb}{\hbox{\reset@font\bfseries ?}%
\ G@refundefinedtrue
\@latex@warning
{Citation `\@citeb' on page \thepage \space undefined}}%
{\@cite@ofmt{\csname b@\@citeb\endcsname}}}}{#1}}
\makeatother

\begin{document}

\title{Coexistence of Pull and Push Communication in Wireless Access for IoT Devices}

 \author{\IEEEauthorblockN{Sara Cavallero\IEEEauthorrefmark{1}, Fabio Saggese\IEEEauthorrefmark{2}, Junya Shiraishi\IEEEauthorrefmark{2}, Shashi Raj Pandey\IEEEauthorrefmark{2}, Chiara Buratti\IEEEauthorrefmark{1}, Petar Popovski\IEEEauthorrefmark{2}\medskip}
\IEEEauthorblockA{
\IEEEauthorrefmark{1}WiLab/CNIT and University of Bologna, Italy.
\IEEEauthorrefmark{2}Department of Electronic System, Aalborg University, Denmark. \\
\IEEEauthorrefmark{1}\{s.cavallero, c.buratti\}@unibo.it, 
\IEEEauthorrefmark{2}\{fasa, jush, srp, petarp\}@es.aau.dk.
\thanks{This work was partly supported by the Villum Investigator Grant ``WATER" from the Velux Foundation, Denmark, partly by the  Horizon Europe SNS ``6G-GOALS'' project with grant 101139232, and partly by the Horizon Europe SNS ``6G-XCEL" project with Grant 101139194.}}}

\maketitle

\begin{abstract}
We consider a setup with \gls{iot}, where a \gls{bs} collects data from nodes that use two different communication modes. The first is \emph{pull-based}, where the \gls{bs} retrieves the data from specific nodes through queries. In addition, the nodes that apply pull-based communication contain a wake-up receiver: upon a query, the \gls{bs} sends \gls{wus} to activate the corresponding \glspl{wud}. The second one is \emph{push-based} communication, in which the nodes decide when to send to the \gls{bs}. Consider a time-slotted model, where the time slots in each frame are shared for both pull-based and push-based communications. Therein, this coexistence scenario gives rise to a new type of problem with fundamental trade-offs in sharing communication resources: the objective to serve a maximum number of queries, within a specified deadline, limits the transmission opportunities for push sensors, and vice versa. This work develops a mathematical model that characterizes these trade-offs, validates them through simulations, and optimizes the frame design to meet the objectives of both the pull- and push-based communications.   
\end{abstract}


\glsresetall

\section{Introduction}
The transition from 5G to 6G communication affects significantly the \gls{iot} domain, where intelligence will play an increasingly large role in the overall system setup, supporting diverse applications~\cite{Survey_5g_IoT}. The increasing intelligence within the communication nodes and devices directs the evolution towards semantic and goal-oriented communication, fostering collaboration and context-aware information sharing~\cite{semantic_comm}. 
Within this \gls{iot} framework, two distinct communication modes take place: \textit{pull-based} communication, where the receiver determines who can send information, leading to scheduled transmissions; and \textit{push-based} communication, where each device autonomously decides when to transmit to the receiver, following the principles of random access. Depending on the specific application requirements, one communication method may be more suitable than the other; however, in a heterogeneous scenario with different types of \gls{iot} devices, these two communication methods can coexist. This coexistence can result in a balanced and adaptive system, optimizing and enhancing the overall performance. 

Only a limited number of studies in the literature have explored their interaction.  In \cite{grandient-opt}, a solution to a distributed convex optimization problem is presented using the push-pull gradient method, involving the push of information about gradients from agents to neighbours and the pull of decisions in the opposite direction. The authors in \cite{talli2024push} proposed an analytical model for optimizing push- and pull-based communication, involving interaction between the actuator and the BS to share their state information. While these works treat the two methods of communication as interdependent, where the action to be performed on the push depends on the result of the pull and vice versa, we approach them as separate processes that must coexist within the same system. 

In our study, we examine a network featuring a 5G-enabled \gls{bs} managing wireless traffic from two device classes: the nodes operating with pull-based communication and those utilizing push-based communication. 
For pull-based communication, this paper considers the system integrating wake-up radio, in which ultra low-power wake-up receiver~\cite{piyare2017ultra} is installed into \gls{iot} devices. The wake-up receiver keeps active monitoring of the \gls{wus} while each node turns off the main radio, saving the wasteful energy consumption of nodes during the idle period. This paper focuses on the well-investigated \gls{idwu}~\cite{piyare2017ultra, 3gpp:38-869}, which enables the \gls{bs} to retrieve the data from the target devices based on its ID. This is suitable for a variety of practical scenarios where the external entity sends a query request to the \gls{bs} through the cloud. 
For push-based communication, devices initiate data transmission with random access to the channel once a data packet is available in their queue.
Therein, pull-based communication prioritizes goal-oriented behaviour to maximize query satisfaction, while push communication aims to maximize the success probability and, consequently, the throughput of push sensors. To the best of our knowledge, this is the first work that exploits the integration of these two communication modes and identifies possible trade-offs between their performance metrics. The performance of this coexistence communication is assessed mathematically, and the validity of the implemented model is evaluated by comparing it with simulations.

\paragraph*{Notation}
$k\sim\mathrm{Poiss}(\mu)$ is Poisson distributed \gls{rv} with mean value $\mu$, i.e., $k\in\mathbb{N}$ has \gls{pmf} $\mc{P}(k, \mu) = \frac{\mu^k  e^{-\mu}}{k!}$; 
$\boldsymbol{B}(m, E)$ denotes the Erlang-B formula with $m$ servers and normalized ingress load $E$. $\lfloor x \rfloor$ denotes the closest integer lower than $x$.

\section{System Model}
\label{sec: system_model}
We consider an \gls{iot} scenario where a \gls{bs} manages access control for two type of devices, or nodes, communicating over a shared wireless link: some nodes are tailored for pull-based communication, while others are for push-based communication. 
Besides receiving \gls{ul} data from various devices within the network, the \gls{bs} also manages queries coming through the cloud, requesting data from specific pull-based \glspl{wud}. 
To simplify the analysis and to show a possible trade-off of the coexistence of pull-push communication, we assume a collision channel and a narrow-band transmission, where a single frequency channel is reserved for both \gls{ul} and \gls{dl} communication\footnote{Remark that this time-frequency structure is equivalent to the one used in 5G \gls{nr} definition, assuming a single \gls{rb} is used~\cite{3gpp:nr}. The extension to the multi-carrier system is left for future studies.}. 

\begin{figure}[t!]
    \centering
    \includegraphics[width=.99\columnwidth]{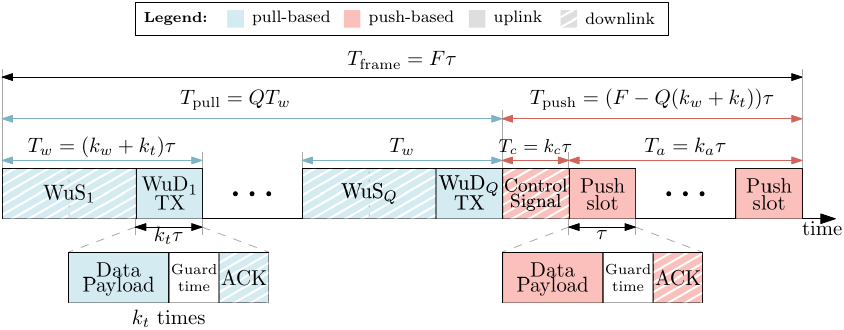}
    \caption{Example of time-frame structure for pull and push communication coexistence, with $Q$ scheduled communications for \glspl{wud}, $k_a$ slots for push-based access, $k_w=$ 2, $k_t=k_c=$ 1.}
    \label{fig:time-diagram}
    \vspace{-5mm}
\end{figure}

As depicted in Fig.~\ref{fig:time-diagram}, the system operates in a time-slotted framework organized into frames, where the duration of each frame, denoted as $T_{\rm frame}$, is split into two segments: $T_{\rm pull}$ and $T_{\rm push}$ dedicated to pull and push-based communication, respectively. Each frame is further divided into $F$ slots, each of duration equal to $\tau$ [s], such that $\tau F = T_{\rm frame}$.
In order to be compliant with the \gls{3gpp} standard~\cite{3gpp:38-912}, $\tau$ is assumed to be a multiple of an \gls{ofdm} symbol, as in~\cite{eucnc, mini-slot}. Moreover, we assume that a single slot has been designed to accommodate an \gls{ul} data transmission, the corresponding \gls{ack} reception, plus an additional guard time to switch from \gls{ul} to \gls{dl} transmissions.
In the following, the details of pull and push communications are given.

\subsection{Pull-based communication}
\label{sec:pull}
Regarding the pull-based communication, we assume that data collection is subject to the reception of specific queries coming through the cloud. These queries contain an identifier (ID) of the device to pull data from (selected depending on the task to be performed), and a deadline that needs to be satisfied, denoted as $L_\mathrm{th}$ [frames]. If a query is not \emph{successfully} served before the deadline, it is \emph{discarded}. 
The queries are coming through the cloud according to a \gls{ppp} with an average arrival rate $\lambda_{q}$ [queries/s]. Accordingly, the number of queries in a frame is 
$n_{q}\sim\mathrm{Poiss}(\overline{n}_q)$ with $\overline{n}_{q} = \lambda_{q} T_{\rm frame}$.
During each frame, the \gls{bs} accumulates the queries, and, in the subsequent frame, it tries to serve them retrieving the data from a subset of \glspl{wud}. 
To retrieve the data from the subsets of target nodes without collisions, this paper applies \gls{idwu}, in which a unique wake-up ID is embedded into a \gls{wus}, allowing the \gls{bs} to collect data from individual nodes based on its own scheduling.
The \glspl{wus} are sent based on the order of queries' arrival at the \gls{bs} -- \gls{fifo} principle. Once the corresponding \gls{wud} has received the \gls{wus}, it switches on its primary radio and transmits its data. Without loss of generality, we assume that $k_w$ slots are needed for the transmission of the \gls{wus} and the activation of the \gls{wud} primary radio, while the \gls{wud} data transmission occupies the subsequent $k_t$ slots.
Therefore, for each \gls{wud}, the scheduled communication has a duration $T_w = (k_w + k_t) \tau$. Consequently, the time dedicated to the pull-based communication is
\begin{equation} \label{eq:tpull}
    T_{\rm pull} = Q T_w = Q (k_w + k_t) \tau,
\end{equation}
where $Q$ represents the number of queries that can be effectively served within that frame. Remark that $Q$ is a fundamental design parameter of the system, representing the amount of resources reserved for pull-based communication.
For simplicity, we assume that each query received in the $t$-th frame must be addressed before the end of the $(t+1)$-th frame, meaning queries have a one-frame deadline ($L_\mathrm{th} = 1$). Hence, if the \gls{bs} cannot fulfill a query for a specific \gls{wud} within the current frame due to insufficient resources, it is discarded.

\subsection{Push-based communication}
\label{sec:push}
The sensors involved in the push-based communication operate intermittently, taking measurements of the environment according to a \gls{ppp} and deciding autonomously whether to transmit the measured data. Accordingly, the generated packets follow a \gls{ppp} with mean arrival rate $\lambda_p$ [packet/s], and the number of packets in a frame $n_p \sim \mathrm{Poiss}(\overline{n}_p)$, with $\overline{n}_p = \lambda_p T_\mathrm{frame}$.

Unlike pull-based communication, 
push-based communication employs contention-based grant-free access, within the portion of the frame $T_\mathrm{push}$, using a Framed-ALOHA scheme. At the beginning of $T_\mathrm{push}$, a subset of $k_{c}$ slots is dedicated to the transmission of a control message -- e.g., a beacon -- from the \gls{bs} to signal readiness to receive push data\footnote{From the \gls{3gpp} standard, this can be seen as similar to the transmission of the \gls{ss} of the RACH procedure~\cite{3gpp:38-912}.}; this control message has a duration of $T_c = k_c \tau$. The remaining time, $T_a = k_{a} \tau$, is available for nodes to access the channel randomly selecting a subset of the $k_{a}$ slots to transmit their queued packets, one per slot. If two or more nodes choose the same slot, a collision occurs, and all the packets transmitted in that slot are lost\footnote{For the sake of simplicity, no re-transmission technique such as \gls{harq} is considered.}.
Accordingly, the time reserved for push communication is $T_{\rm push} = (k_c + k_a) \tau$ which can be re-written as a function of $Q$, following 
\begin{equation} \label{eq:tpush}
    T_{\rm push} = T_{\rm frame} - T_{\rm pull} = (F - Q(k_w + k_t)) \tau.
\end{equation}

\subsection{Pull and push trade-off}
In the proposed scenario, we need to manage the finite time resources of the \gls{mac} frame shared between pull and push nodes taking into account the different requirements of the two traffic: pull-based communication aims to maximize the average number of successfully served queries, while push-based communication wants to maximize the throughput. 
To reach these two objectives, it is necessary to find a trade-off on the time resources allocated for the two communications, considering that $T_\mathrm{frame}$ remains fixed. 
A high number of queries collected in the previous frame requires reserving more slots for pull communication to wake up all the intended \glspl{wud} and ensure their transmissions. In this case, it becomes essential to increase $T_\mathrm{pull}$, which simultaneously decreases $T_\mathrm{push}$. In turn, this increases the collision probability for \gls{ul} push transmissions, because of the lower number of available slots, leading to a lower throughput of the push devices. 
These communication requirements can be translated into a unique metric, accounting for both, the \emph{success probability} of serving the queries and the \emph{success probability} of push access (see Section~\ref{sec: metrics}). 
The goal is to determine the values of $T_\mathrm{pull}$ and $T_\mathrm{push}$ able to achieve a trade-off for the communication coexistence. Since $T_\mathrm{frame} = T_\mathrm{pull} + T_\mathrm{push}$, and both terms depends on the number of queries that can be served (see eq.~\eqref{eq:tpull} and~\eqref{eq:tpush}), our goal translates into finding the values of $Q$ striking a trade-off between success probability of queries and push packets, taking into account that $k_a =  F - k_c - Q\, (k_w + k_t)$ and $Q = \left\lfloor \frac{F - k_c - k_a}{k_w + k_t} \right\rfloor$.

\section{Performance Evaluation}
\label{sec: metrics}
Throughout this section, we aim to relate the metrics of pull- and push-based communications with the value of $Q$ controlling the portion of time reserved to the two traffic.

\subsection{Success probability of serving the queries}
In pull-based communication, the \gls{bs} must timely collect fresh data from the \gls{wud}, meeting a single-frame deadline. If the \gls{bs} cannot obtain data from a specific \gls{wud} by the deadline, it discards the query request. This occurs when the available resources for pull-based communication, denoted as $Q$, are insufficient to handle the number of queries received in the previous frame.

To analyze the success probability of serving the queries, we remark that the current setting is equivalent to a multi-server queuing model with null queuing capacity. Following Kendall's notation, this can be represented by a $M/D/Q/0$ queue model, according to the fact that only the first $Q$ queries can be served, and the others are discarded. Under this model, the Erlang-B formula represents the discarding probability when the number of sources is infinite~\cite{Thomas_1976, erlang_paper}. Hence, the success probability of serving the queries is approximated as:
\vspace{-2mm}
\begin{equation}
  p_{s}^{(q)} \approx 1- \boldsymbol{B}(Q, \overline{n}_{q}) = 1- \frac{\frac{\overline{n}_q^{\,Q}}{Q!}}{\sum_{i=0}^{Q}{\frac{\overline{n}_q^{\, i}}{i!}}}.
  \label{eq:p_s_pull}
\end{equation}
The approximation becomes exact when 
the average number of queries $\overline{n}_q$ (pull sources), grows to infinity.

From this metric, it is possible to find the average number of successfully served queries.
Since each query has a success probability independent from query to query, and from the thinning property of \gls{ppp}~\cite{Baddeley2007}, the number of successfully served queries, $n_s = n_q \, p_{s}^{(q)}$, is still Poissonian, with mean:
\vspace{-2mm}
\begin{equation}
  \overline{n}_{s} = \sum_{n_q=0}^{\infty} n_{q} \;  p_{s}^{(q)} \; \mathcal{P}(n_{q}, \overline{n}_{q}) = \overline{n}_{q} \; p_{s}^{(q)}.
  \label{eq:N_s}
\end{equation}

\subsection{Average Throughput for push-based communication}
For the push-based communication, we are interested in analyzing the average throughput, $S_{\rm push}$ [packets/s], defined as the average number of packets per unit of time successfully received by the \gls{bs} from the push devices. Consequently, $S_{\rm push}$ is related to the success probability of push access $p_s^{(p)}$.

In order to formalize the latter, let us denote the probability of successful access in a frame conditioned to the knowledge of the number of transmitted push packets $n_p$ as $p_{s | n_p}^{(p)}$. Under the Framed ALOHA scheme and collision channel assumptions, $p_{s | n_p}^{(p)}$ depends on the available slots for the push $k_a$, with the assumption that at least one slot is available for push access, i.e., $k_a > 0$.
If $k_a= 1$, the probability of access is 1 if $n_p \le 1$, and 0 otherwise, due to the unavoidable collisions; if $k_a \ge 1$, the probability of accessing the channel is exactly $1$ with $n_p = 0$, because no collisions can occur, while it is $(1 - 1/k_a)^{n_p -1}$ for $n_p \ge 1$, representing the case that only a single packet selects a slot. Hence, we can write:
\begin{equation}
\begin{split}
p_{s | n_p}^{(p)} \hspace{-2mm}= \hspace{-1mm}
\begin{cases}
    0, 
    &\text{if } k_a = 1, n_p > 1, \\
    \left( 1 - \frac{1}{k_a} \right)^{n_p - 1}, 
    &\text{if } k_a > 1, n_p \ge 1,    \\
    1, &\text{if } 
    \begin{cases}
        k_a = 1, \\
        n_p \le 1,
    \end{cases} \hspace{-3mm}\text{or }
    \begin{cases}
        k_a > 1, \\
        n_p = 0.
    \end{cases}
\end{cases}
    \label{eq:pmac_np}
\end{split}
\end{equation}


From~\eqref{eq:pmac_np}, it is possible to compute the success probability of push access considering two different cases: $k_a = 1$ and $k_a >1$. If $k_a = 1$, the access probability is equal to the probability that 0 or 1 packets are transmitted, i.e., $p_s^{(p)} = \mathcal{P}(0, \overline{n}_p) + \mathcal{P}(1, \overline{n}_p)$. If $k_a > 1$, the probability of access can be obtained by applying the law of total probability, i.e., $p_{s}^{(p)} = \sum_{n_p = 0}^{\infty} {p_{s | n_p}^{(p)} \, \mathcal{P}(n_p, \overline{n}_p)} = \mathcal{P}(0, \overline{n}_p) + \sum_{n_p = 1}^{\infty} {\left( 1 - \frac{1}{k_a} \right)^{n_p - 1}\mathcal{P}(n_p, \overline{n}_p)}$. This results in:
\begin{equation}
\begin{split}
p_{s}^{(p)} =
\begin{cases}
    (1 + \overline{n}_p) \, e^{-\overline{n}_p}, &\text{if } k_a = 1, \\
    e^{-\overline{n}_p} \, \frac{k_a \, e^{\frac{(k_a - 1)\, \overline{n}_p}{k_a}}\, - 1}{k_a - 1},  &\text{if } k_a > 1.
    \label{eq:p_macReal}
\end{cases}
\end{split}
\end{equation}

Similarly to the previous case, based on~\eqref{eq:pmac_np}, we can derive the average throughput for push nodes as in \cite{S_csma, S_uav}, obtaining
\vspace{-2.2mm}
\begin{equation}
\begin{split}
 S_{\rm push} &= \frac{1}{T_{\rm frame}} \hspace{-1.1mm} \sum_{n_p=0}^{\infty}\hspace{-1.1mm} n_p \;  p_{s | n_p}^{(p)} \mathcal{P}(n_p, \overline{n}_p)
  = \frac{\overline{n}_p}{T_{\rm frame}} \, e^{- \,\frac{\overline{n}_p}{k_a}}. 
  \label{eq:S_final}
  \end{split}
\end{equation}

\subsection{Success probability trade-off}
Instead of jointly maximizing both pull and push success probabilities, we characterize a performance trade-off defining the weighed average success probability for the communication coexistence as 
\begin{equation}
    p_{s} = w_q\, p_{s}^{(q)} + w_p \, p_{s}^{(p)},\label{eq:pmac}
\end{equation}
where $w_q~\in~[0, 1]$ and $w_{p}\in~[0, 1]$ are positive weights useful to target requirements of pull- and push-based communications, respectively. 
Since both $p_{s}^{(q)}$ and $p_{s}^{(p)}$ are functions of $Q$, it is possible to find the value of $Q$ that maximizes $p_{s}$, under a specific set of weights and system settings.
As a show case, we select $w_q=\frac{\lambda_q}{\lambda_q + \lambda_p}$ and $w_p=\frac{\lambda_p}{\lambda_q + \lambda_p}$, considering the fairness in terms of traffic load for both pull and push nodes.

\section{Numerical Results}
\label{sec: results}
In this section, we numerically benchmark the performance of the proposed system. 
We build a custom simulator that implements both the pull and the push-based communications, according to the system model described in Section~\ref{sec: system_model}. Each simulation lasts $10^{5}\; T_{\rm frame}$; the averaged values are plotted in the following figures. Unless stated otherwise, Table~\ref{tab:system_parameter_settings} reports the parameter values used for numerical results.

\def\arraystretch{1.3}
\begin{table}[t!]
\centering
\footnotesize
\caption{Simulation parameters.}
\label{tab:system_parameter_settings}
\begin{tabular}{l|l|l}
\hline
\textbf{Parameter} & \textbf{Description} & \textbf{Value}  \\\hline
$\tau$ & Slot duration [s] & 0.25 ms\tablefootnote{According to 3GPP concept of mini-slots~\cite{3gpp:38-912}, we assume a slot$= 7$ \gls{ofdm} symbols and a sub-carrier spacing $\Delta_f = 30$ kHz, working in FR1.} \\ 
$F$ & Slots in a frame & 101 \\ 
$T_{\rm frame}$ & Frame duration [s] &  25.25 ms \\ 
$k_w$ & Slots for \gls{wus} & 4\tablefootnote{Value set according to 3GPP \cite[Table 7.1.2.2-5]{3gpp:38-869}.}\\ 
$k_t$ & Slots in for \gls{wud} TX & 1 \\
$k_c$ & Slots for control part & 1 \\ 
$k_a$ & Slots for Framed ALOHA TX & 1 \\ 
\end{tabular}
\end{table}
\label{sub:params}

\subsection{Validation of the analytical model}\label{sec:Num_basic_trade_off}

\begin{figure}[t!]
\centering
    \input{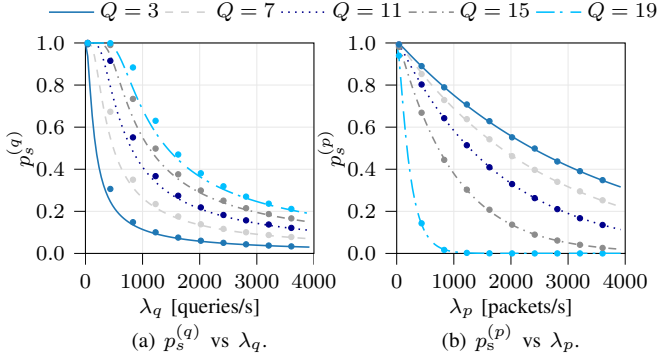}
    \vspace{-0.5cm}
    \caption{Validation of analytical models with different $Q$ values.}
    \label{fig:analytical}
    \vspace{-5mm}
\end{figure}

Fig.~\ref{fig:pq_fig} shows the success probability of serving the queries, $p_s^{(q)}$ versus the average arrival rate of the queries, $\lambda_q$, while Fig.~\ref{fig:pmac_fig} depicts the success probability $p_s^{(p)}$ against the average arrival rate for the push-based communication, $\lambda_p$. These results are obtained from both our theoretical analysis (lines) and computer simulation (circular symbols) considering different $Q$ values. 
Fig.~\ref{fig:pmac_fig} shows a perfect correspondence between the theoretical analysis and the numerical simulation, validating our results. 
In Fig.~\ref{fig:pq_fig}, there is a slight difference between the analytical and simulation results for a range of low values of $\lambda_q$, due to the infinite population of sources assumption of the Erlang-B formula. Nevertheless, eq.~\eqref{eq:p_s_pull} is a tight lower bound of the actual success probability $\forall \lambda_q$, justifying the use of our analytical model to characterize the system performance.

Finally, the results already show that a fundamental trade-off on time resources exists: the higher is $Q$, the greater $p_s^{(q)}$, whereas the lower the $p_{s}^{(p)}$, and vice versa.
This result indicates the importance of the choice of the parameter $Q$ considering both types of traffic.

\subsection{Design guidelines for coexistence of pull and push-based communication}
\begin{figure}[tb]
\centering
    \input{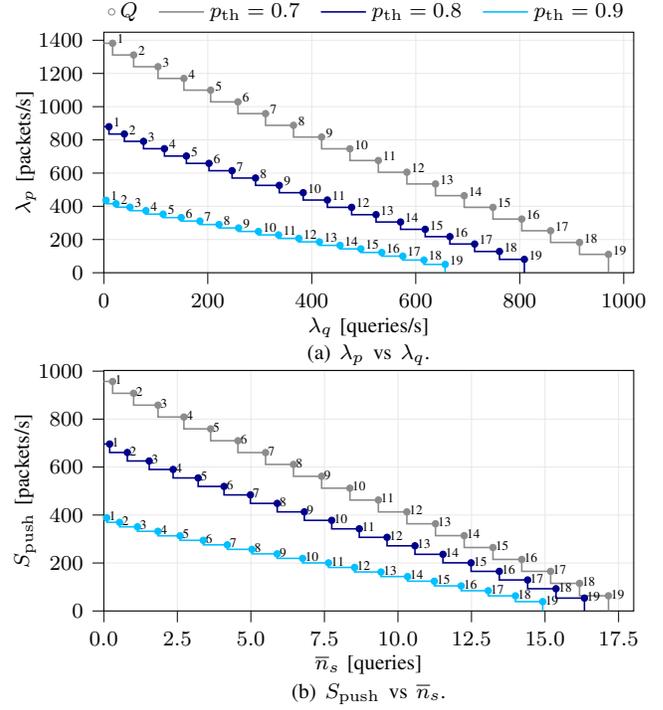}
    \caption{Design guidelines for pull and push communications when $p_s^{(q)} = p_s^{(p)} = p_{\rm th}$.}
    \label{fig:design_guidelines}
    \vspace{-5mm}
\end{figure}
Fig.~\ref{fig:design_guidelines} shows the achievable performance of the system constrained to a target success probability for pull and push traffic.
To obtain the plot, we set a requirement on the success probabilities: $p_s^{(q)} = p_s^{(p)} = p_{\rm th}$.  For each $Q$ value, we compute the maximum arrival rates $\lambda_q^{\rm (max)}$ and $\lambda_p^{\rm (max)}$ that can be supported satisfying this constraint.
%
Knowing the arrival rates, for each $Q$ value, we determine the maximum number of successfully served queries $\overline{n}_s (\lambda_q^{(\max)})$ and the maximum  throughput $S_{\rm push} (\lambda_p^{(\max)})$ using \eqref{eq:N_s}, and \eqref{eq:S_final}, respectively. 

Fig.~\ref{fig:lambdaP_vs_lambdaQ} displays the acceptable traffic $\lambda_p$ vs $\lambda_q$ for $p_{\rm th} = [0.7; 0.8; 0.9]$. Increasing $Q$ generally leads to a decrease in the acceptable $\lambda_p$, as $\lambda_q$ rises, accommodating higher pull traffic to meet the $p_{\rm th}$ requirement. Lower $p_{\rm th}$ allows for higher acceptable arrival rates.
Fig.~\ref{fig:Spush_vs_ns} shows the throughput $S_{\rm push}$ as a function of $\overline{n}_s$ for the same $p_{\rm th}$ values.
The behaviour of this graph aligns with the previous one, where increasing $Q$ and/or decreasing $p_{\rm th}$, directly correlates with an increase in $\lambda_q$ and therefore a higher number of satisfied queries $\overline{n}_s$. This corresponds with a simultaneous decrease in the supported input traffic for push communication, resulting in a lower achieved throughput.
For example, considering $p_{\rm th} = 0.8$ and $Q = 2$, the success requirement is met up to $\lambda_q = 39$ queries/s, and $\lambda_p = 835$ packets/s. This results in an achievable throughput $S_{\rm push} = 660$ packets/s and an average number of successfully served queries $\overline{n}_s = 0.5$. If $\lambda_q$ increases, raising $Q$ to 3 becomes necessary, leading to a decrease in input push traffic to $\lambda_p = 791$ packets/s corresponding to $S_{\rm push} = 625$ packets/s. The region under each curve indicates the set of achievable performance for push and pull at each $Q$ value, facilitating the selection of suitable parameters based on specific requirements.

\subsection{Trade-off pull/push-based communication}

\begin{figure}[t!]
\begin{center}
\begin{tikzpicture}

\begin{axis}[
height=2.8cm,
legend columns=-1,
legend cell align={left},
legend style={at={(0.5, 1,05)}, anchor=south, draw=none, /tikz/every even column/.append style={column sep=0.5cm}},
xlabel={$Q$},
xmin=0, xmax=20,
ylabel={\(\displaystyle p_\mathrm{s}\)},
ymajorgrids,
ymin=0, ymax=0.9,
ytick style={color=black},
ytick={0,0.1,0.2,0.3,0.4,0.5,0.6,0.7,0.8,0.9},
yticklabels={
  \(\displaystyle {0.0}\),
  \(\displaystyle {0.1}\),
  \(\displaystyle {0.2}\),
  \(\displaystyle {0.3}\),
  \(\displaystyle {0.4}\),
  \(\displaystyle {0.5}\),
  \(\displaystyle {0.6}\),
  \(\displaystyle {0.7}\),
  \(\displaystyle {0.8}\),
  \(\displaystyle {0.9}\)
},
legend image code/.code={
        \draw [#1] (0cm,-0.1cm) rectangle (0.15cm,4pt); },
xlabel shift=-3pt,
]
\addlegendimage{draw=lightskyblue,fill=lightskyblue}
\addlegendentry{$\lambda_q = 0.5\lambda_p$}
\addlegendimage{draw=gray,fill=gray}
\addlegendentry{$\lambda_q = \lambda_p$}
\addlegendimage{draw=darkblue,fill=darkblue}
\addlegendentry{$\lambda_q = 1.5\lambda_p$}

\draw[draw=lightskyblue,fill=lightskyblue] (axis cs:0.7,0) rectangle (axis cs:0.9,0.635498482273987);
\draw[draw=lightskyblue,fill=lightskyblue] (axis cs:1.7,0) rectangle (axis cs:1.9,0.675415842898733);
\draw[draw=lightskyblue,fill=lightskyblue] (axis cs:2.7,0) rectangle (axis cs:2.9,0.712764048571907);
\draw[draw=lightskyblue,fill=lightskyblue] (axis cs:3.7,0) rectangle (axis cs:3.9,0.746907830177465);
\draw[draw=lightskyblue,fill=lightskyblue] (axis cs:4.7,0) rectangle (axis cs:4.9,0.777104890058501);
\draw[draw=lightskyblue,fill=lightskyblue] (axis cs:5.7,0) rectangle (axis cs:5.9,0.802547592327496);
\draw[draw=lightskyblue,fill=lightskyblue] (axis cs:6.7,0) rectangle (axis cs:6.9,0.822450441174929);
\draw[draw=lightskyblue,fill=lightskyblue] (axis cs:7.7,0) rectangle (axis cs:7.9,0.836179318516131);
\draw[draw=lightskyblue,fill=lightskyblue] (axis cs:8.7,0) rectangle (axis cs:8.9,0.843384157401843);
\draw[draw=lightskyblue,fill=lightskyblue] (axis cs:9.7,0) rectangle (axis cs:9.9,0.844061323616677);
\draw[draw=lightskyblue,fill=lightskyblue] (axis cs:10.7,0) rectangle (axis cs:10.9,0.838472525828132);
\draw[draw=lightskyblue,fill=lightskyblue] (axis cs:11.7,0) rectangle (axis cs:11.9,0.826904095902844);
\draw[draw=lightskyblue,fill=lightskyblue] (axis cs:12.7,0) rectangle (axis cs:12.9,0.809321916056759);
\draw[draw=lightskyblue,fill=lightskyblue] (axis cs:13.7,0) rectangle (axis cs:13.9,0.784982906658994);
\draw[draw=lightskyblue,fill=lightskyblue] (axis cs:14.7,0) rectangle (axis cs:14.9,0.751968438375002);
\draw[draw=lightskyblue,fill=lightskyblue] (axis cs:15.7,0) rectangle (axis cs:15.9,0.706431246424316);
\draw[draw=lightskyblue,fill=lightskyblue] (axis cs:16.7,0) rectangle (axis cs:16.9,0.641116121421828);
\draw[draw=lightskyblue,fill=lightskyblue] (axis cs:17.7,0) rectangle (axis cs:17.9,0.542898637959062);
\draw[draw=lightskyblue,fill=lightskyblue] (axis cs:18.7,0) rectangle (axis cs:18.9,0.400040100178558);

\draw[draw=gray,fill=gray] (axis cs:0.9,0) rectangle (axis cs:1.1,0.479133075223878);
\draw[draw=gray,fill=gray] (axis cs:1.9,0) rectangle (axis cs:2.1,0.512442306090227);
\draw[draw=gray,fill=gray] (axis cs:2.9,0) rectangle (axis cs:3.1,0.544957948235908);
\draw[draw=gray,fill=gray] (axis cs:3.9,0) rectangle (axis cs:4.1,0.576535560428026);
\draw[draw=gray,fill=gray] (axis cs:4.9,0) rectangle (axis cs:5.1,0.606998312724188);
\draw[draw=gray,fill=gray] (axis cs:5.9,0) rectangle (axis cs:6.1,0.636128989349349);
\draw[draw=gray,fill=gray] (axis cs:6.9,0) rectangle (axis cs:7.1,0.663660012808812);
\draw[draw=gray,fill=gray] (axis cs:7.9,0) rectangle (axis cs:8.1,0.689260982252161);
\draw[draw=gray,fill=gray] (axis cs:8.9,0) rectangle (axis cs:9.1,0.71252294535178);
\draw[draw=gray,fill=gray] (axis cs:9.9,0) rectangle (axis cs:10.1,0.73293797867269);
\draw[draw=gray,fill=gray] (axis cs:10.9,0) rectangle (axis cs:11.1,0.749871123297304);
\draw[draw=gray,fill=gray] (axis cs:11.9,0) rectangle (axis cs:12.1,0.762518189044);
\draw[draw=gray,fill=gray] (axis cs:12.9,0) rectangle (axis cs:13.1,0.769835013785166);
\draw[draw=gray,fill=gray] (axis cs:13.9,0) rectangle (axis cs:14.1,0.770406351502902);
\draw[draw=gray,fill=gray] (axis cs:14.9,0) rectangle (axis cs:15.1,0.762184588422971);
\draw[draw=gray,fill=gray] (axis cs:15.9,0) rectangle (axis cs:16.1,0.741946621726327);
\draw[draw=gray,fill=gray] (axis cs:16.9,0) rectangle (axis cs:17.1,0.704165648838129);
\draw[draw=gray,fill=gray] (axis cs:17.9,0) rectangle (axis cs:18.1,0.639126194549173);
\draw[draw=gray,fill=gray] (axis cs:18.9,0) rectangle (axis cs:19.1,0.538313282542248);

\draw[draw=darkblue,fill=darkblue] (axis cs:1.1,0) rectangle (axis cs:1.3,0.384042705901112);
\draw[draw=darkblue,fill=darkblue] (axis cs:2.1,0) rectangle (axis cs:2.3,0.411589892915833);
\draw[draw=darkblue,fill=darkblue] (axis cs:3.1,0) rectangle (axis cs:3.3,0.438703282107409);
\draw[draw=darkblue,fill=darkblue] (axis cs:4.1,0) rectangle (axis cs:4.3,0.465313742971461);
\draw[draw=darkblue,fill=darkblue] (axis cs:5.1,0) rectangle (axis cs:5.3,0.491337031661934);
\draw[draw=darkblue,fill=darkblue] (axis cs:6.1,0) rectangle (axis cs:6.3,0.516669436371726);
\draw[draw=darkblue,fill=darkblue] (axis cs:7.1,0) rectangle (axis cs:7.3,0.541181788980235);
\draw[draw=darkblue,fill=darkblue] (axis cs:8.1,0) rectangle (axis cs:8.3,0.564711049831965);
\draw[draw=darkblue,fill=darkblue] (axis cs:9.1,0) rectangle (axis cs:9.3,0.587048181955461);
\draw[draw=darkblue,fill=darkblue] (axis cs:10.1,0) rectangle (axis cs:10.3,0.607920163782624);
\draw[draw=darkblue,fill=darkblue] (axis cs:11.1,0) rectangle (axis cs:11.3,0.626962395792078);
\draw[draw=darkblue,fill=darkblue] (axis cs:12.1,0) rectangle (axis cs:12.3,0.643674702593193);
\draw[draw=darkblue,fill=darkblue] (axis cs:13.1,0) rectangle (axis cs:13.3,0.657348010260379);
\draw[draw=darkblue,fill=darkblue] (axis cs:14.1,0) rectangle (axis cs:14.3,0.666935924116661);
\draw[draw=darkblue,fill=darkblue] (axis cs:15.1,0) rectangle (axis cs:15.3,0.670817297264836);
\draw[draw=darkblue,fill=darkblue] (axis cs:16.1,0) rectangle (axis cs:16.3,0.666333602146122);
\draw[draw=darkblue,fill=darkblue] (axis cs:17.1,0) rectangle (axis cs:17.3,0.648866677017237);
\draw[draw=darkblue,fill=darkblue] (axis cs:18.1,0) rectangle (axis cs:18.3,0.610328488085236);
\draw[draw=darkblue,fill=darkblue] (axis cs:19.1,0) rectangle (axis cs:19.3,0.543493393180213);
\end{axis}

\end{tikzpicture}
    \vspace{-3mm}
    \caption{$p_{s}$ vs $Q$ for $\lambda_p$ = 500 packets/s and different $\lambda_q$ values.}
    \label{fig:pmac_Q}
\end{center}
\vspace{-6mm}
\end{figure}
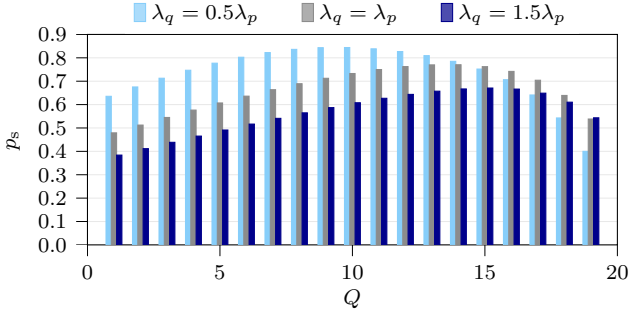

Fig.~\ref{fig:pmac_Q} shows the behaviour of the weighted average success probability $p_{s}$ of eq.~\eqref{eq:pmac} as a function of $Q$ for $\lambda_p = 500$ packets/s and $\lambda_q = [0.5; 1; 1.5] \lambda_p$. 
Results show that the highest $p_{s}$ is obtained with $Q =[10; 14; 15]$ when $\lambda_q = [0.5; 1; 1.5] \lambda_p$, respectively. As expected, increasing $\lambda_q$, the maximum of $p_{s}$ is achievable increasing $Q$, meaning that more slots dedicated to the pull-based communication are needed.

\begin{figure}[t!]
\begin{center}
    \input{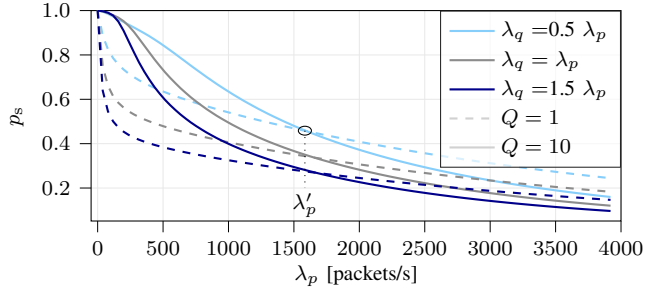}
    \vspace{-7mm}
    \caption{$p_{s}$ vs $\lambda_p$ for different $\lambda_q$ (color coded) and $Q$ values (solid and dashed lines).}
    \label{fig:pmac_lambdaP}
\end{center}
\vspace{-6mm}
\end{figure}

Finally, in Fig.~\ref{fig:pmac_lambdaP}, we evaluate $p_{s}$ as a function of $\lambda_p$ for $\lambda_q = [0.5; 1; 1.5] \lambda_p$ and $Q \in\{1,10\}$. This graph shows that $p_{s}$ is a monotonically decreasing function of both  $\lambda_p$ and $\lambda_q$. Notably, when $\lambda_q$ is held constant, the curves for $Q = 1$ and $Q = 10$ intersect at a specific arrival rate $\lambda_p'$. When $\lambda_p < \lambda_p'$, it is advisable to maintain $Q = 10$, allocating more slots for pull communication. Conversely, for $\lambda_p > \lambda_p'$, reducing $Q$ to 1 is preferable. In this scenario, the impact of push communication traffic is more significant on system performance, and allocating more slots to push results in an enhancement of overall performance. For clarity, the graph shows only two $Q$ values, but it can be extended to all $Q$ values, facilitating the identification of the optimal solution based on the system's input load ($\lambda_p$ and $\lambda_q$).

\section{Conclusions}
\label{sec: conclusion}
This paper studies the integration of pull-enabled \glspl{wud} and intermittently active push sensors operating within the same time frame structure in an \gls{iot} setting. 
A mathematical model, validated through simulations, highlights the interdependence of both communications and the need of adjusting the time allocation reserved for pull and push to achieve optimal performance. 
Based on the knowledge of the incoming push and pull traffic, the time allocation that optimally balances the overall performance in a single frame has been found. Future works will investigate the optimal performance considering a longer time horizon, as well as learning that is capable to capture the correlations among the traffic arrivals and data. 


\footnotesize
\bibliographystyle{IEEEtranNoURL}
\bibliography{bibl}

\end{document}